\documentstyle[12pt]{article}
\begin{document}
\title{The Fuzzy Space Time Paradigm}
\author{B.G. Sidharth\\
International Institute of Applicable Mathematics \& Information Sciences\\
B.M. Birla Science Centre, Adarshnagar, Hyderabad - 500 063,
India}
\date{}
\maketitle
\begin{abstract}
We briefly review theories of fuzzy spacetime or with a spacetime cut off, particularising on the author's own model which correctloy predicted the present cosmological model of an accelerating universe and dark energy, and which points to a unification of electromagnetism and gravitation and which has several ramifications including the prediction of deviations from Special Relativity at Ultra High Energies, as observations of Ultra High Energy Cosmic Rays seem to confirm, amongst other issues.
\end{abstract}
\section{Introduction}
It is well known that Quantum Theory and General Relativity have been called the twin pillars of Physics of the twentieth Century. On the one hand, while each in its own right explained aspects of the universe to a certain extent, there are still many unanswered questions. For example space-time singularities, termed by John Wheeler as the Greatest Crisis of Physics, some eighteen arbitrary parameters in the standard model, elusive monopoles, gravitational waves and dark matter and so on.\\
On the other hand, both these fields have resisted a unified description, inspite of several decades of effort. Such a unified description has rightly been called the Holy Grail of Physics, but appears to be a distant goal even today.\\
After fruitless decades, it would be reasonable to explore alternative approaches. Such a revolution is under way. It must be observed that, be it Quantum Field Theory and the standard model or General Relativity, the space time used is a differentiable manifold-this is the common platform on which the two irreconcible pillars stand. Perhaps this is an approximation? Infact Mandelbroit's work on Fractals has clearly brought out that the smooth curves of Classical Mathematics are to be replaced in real life, by fractal structures, previously dismissed as pathological cases.\\
Indeed a few current approaches, already share this philosophy. For example, Quantum Gravity and Quantum SuperString Theory. Here, there is a minimum space-time interval, the Planck scale, which breaks the point space-time of differentiable manifolds. However both these new approaches, though promising, operate at energy scales, which are beyond present day verification. For this and other reasons, a few other approaches are being considered, which rely on an underlying non-commutative geometrical structure of space-time, as do the above two approaches. For example the approach of Connes, Madore and others using the apparatus of non-commutative geometry.\\
We on the other hand consider a similar approach, but this is a more physical model. For example the Kerr-Newman Black Hole of Classical Physics and General Relativity describes the electron's purely Quantum Mechanical g=2 factor. But the price one has to pay is the naked singularity, or equivalently, the complex space coordinate. Curiously enough, the space coordinate of the Dirac electron has precisely the same non Hermitian or complex coordinate. In Quantum Theory, this is due to zitterbewegung effects, which are eliminated, as Dirac pointed out, by averaging out the Compton scale: Space-time points have no physical meaning. Compton scale intervals, complex coordinates, spin and non-commutative geometry, are all symptomatic or indicative of the underlying fuzzy spacetime. It must be stressed that the spacetime of Classical Physics was either Galilean in which space and time were distinct (as brought out by the absolute status enjoyed by the concept of simultaneity) or Lorentzian which was a modification of Galilean spacetime with the introduction of the physical postulates of Special Relativity, or finally it was Reimannian as in General Relativity. Quantum Field Theory also rests on Lorentzian spacetime. In all these cases, the concept is Newtonian, in that spacetime was a container or stage within which the actors of matter, energy and interactions played their parts, even modifying the stage. However, the new concept of spacetime is Liebnitzian, in that, the actors create or define the stage itself.  It is now possible to circumvent space-time singularities and even divergences.\\
This approach gives rise to a Cosmology, which correctly predicted an ever expanding and accelerating universe, as also dark energy at the expense of dark matter. This also paves the way for reconciliation between General Relativity and Quantum Theory. One way to understand this is to recall Wheeler's analysis of the divide-the absence of spin half in General Relativity and curvature in Quantum Theory. Both these are achieved and a Weyl like theory results. These ideas are in harmony with recent observations, which indicate a neutrino mass and time varying "constants" of Physics.\\
The laws of the universe are "thermodynamic", rather than rigid and iron cast.
\section{Further Considerations}
To see all this in greater detail, we observed that if we treat an electron as a Kerr-Newman Black Hole, then even though we get the correct Quantum Mechanical $g=2$ factor, the horizon of the Black Hole becomes complex \cite{cu,mwt}.
\begin{equation}
r_+ = \frac{GM}{c^2} + \imath b, b \equiv (\frac{G^2Q^2}{c^8} + a^2 - \frac{G^2M^2}{c^4})^{1/2}\label{e1}
\end{equation}
$G$ being the gravitational constant, $M$ the mass and $a \equiv L/Mc,L$ being the angular momentum. While (\ref{e1}) exhibits a naked singularity, and as such has no physical meaning, the position coordinate for a Dirac particle is given by
\begin{equation}
x = (c^2p_1H^{-1}t) + \frac{\imath}{2} c\hbar (\alpha_1 - cp_1H^{-1})H^{-1}\label{e2}
\end{equation}
an expression that is very similar to (\ref{e1}). Infact the imaginary parts of both (\ref{e1}) and (\ref{e2}) are the same, being of the order of the Compton wavelength.\\
It is at this stage that a proper physical interpretation begins to emerge. Dirac himself observed that to interpret (\ref{e2}) meaningfully, it must be remembered that Quantum Mechanical measurements are really averaged over the Compton scale: Within the scale there are the unphysical zitterbewegung effects.\\
Once such a minimum space time scale is invoked, then we have a non commutative geometry as shown by Snyder some fifty years ago:
$$[x,y] = (\imath a^2/\hbar )L_z, [t,x] = (\imath a^2/\hbar c)M_x, etc.$$
\begin{equation}
[x,p_x] = \imath \hbar [1 + (a/\hbar )^2 p^2_x];\label{e3}
\end{equation}
The relations (\ref{e3}) are compatible with Special Relativity. Indeed such minimum space time models were studied for several decades, precisely to overcome the divergences encountered in Quantum Field Theory.\\
Before proceeding further, it may be remarked that when the square of the Compton wavelength can be neglected, then we return to point Quantum Theory.\\
It is interesting that starting from the Dirac coordinate in (\ref{e2}), we can deduce the non commutative geometry (\ref{e3}), independently. This is because when we generalise the complex coordinate (\ref{e2}) to three dimensions, then ($\imath , 1$ goes over to $\vec \sigma , I$), where $\sigma$s are the Pauli matrices. We at once deduce spin and Special Relativity and the geometry (\ref{e3}). This is a transition that has been long overlooked \cite{bgsfpl}. Conversely it must be mentioned that spin half itself is relational and refers to three dimensions.\\
Equally interesting is the fact that starting from the geometry (\ref{e3}) we can deduce the Dirac equation itself. Infact if we consider a time shift transformation of the wave function
$$| \psi' > = U (R)| \psi >$$
we get,
$$\psi' (x_j) = [1 + \imath \epsilon (\imath x_j \frac{\partial}{\partial x_j}) + 0 (\epsilon^2)] \psi (x_j)$$
on using the Lorentz covaraint relations (\ref{e3}), which has been shown to lead to the Dirac equation, $\epsilon$ being the Compton time \cite{cu}.\\
We started with the Kerr-Newman Black Hole. Infact the derivation of the Kerr-Newman Black Hole itselof begins with a complex shift, which Newman has found inexplicable even after several decades \cite{newman}. The unanswered question has been, why does a complex shift somehow represent spin about that axis? The answer to this question lies in the above considerations. Complexified space time is symptomatic of fuzzy space time and a non commutative geometry and Quantum Mechanical spin \cite{bgscst}. Indeed Zakrzewsky has shown in a classical context that non commutativity implies spin.
\section{The Compton wavelength and Phase Transition Consideration}
In earlier work, it was shown how one could consider elementary particles as forming out of a background Zero Point Field echoing the ideas of Einstein that particles were condensates of an electromagnetic field \cite{ijmpa}. The starting point was a Schrodinger like equation, allowing non local amplitudes within the Compton wavelength:
\begin{equation}
\imath \hbar \frac{\partial \psi}{\partial t} = \frac{-\hbar^2}{2m'} \frac{\partial^2 \psi}{\partial x^2} + \int \psi^* (x') \psi (x) \psi (x') U(x') dx',\label{e4}
\end{equation}
The integralk in (\ref{e4}) is over a Compton wavelength outside which the function $U$ vanishes, this function being introduced merely for convenience. (\ref{e4}) then becomes identical to the Landau-Ginzburg equation,
\begin{equation}
- \frac{\hbar^2}{2m} \nabla^2 \psi + \beta |\psi |^2 \psi = -\alpha \psi\label{e5}
\end{equation}
The coherence length of the Landau-Ginzburg-Schrodinger equation (\ref{e5}) turns out to be precisely the Compton wavelength \cite{newcosmos}. In other words the elementary particles are created in a phase transition type of a phenomena, resembling inflationary scenarios. One could equivalently look upon the creation of particles in a fluid mechanics context as the formation of Benard cells \cite{bgsnature}. In this case we could compare the Schrodinger equation with the Navier-Stokes equation and the Compton wavelength condition becomes identical to the critical value of the Quantized vortex length.\\
It must be mentioned that in the above considerations, there is a ZPF dominated pre space time which is similar to the uniform fluid near the critical point. This is pre space time because, the total homogeneity does not allow a meaningful definition of coordinates or a metric. It is only at the critical point when Benard cells or elementary particles are formed that space time becomes meaningful in a relational sense (Cf. also \cite{Prig}).
\section{The Unification of Gravitation and Electromagnetism}
The identification of the Kerr-Newman Black Hole of classical physics with the Quantum Mechanical electron already points to a unified description of gravitation and electromagnetism. This can be seen directly from the non commutative geometry \cite{annales}. Indeed let us start with the expression for the metric 
\begin{equation}
ds^2 = g_{\mu \nu} dx^\mu dx^\nu\label{e6}
\end{equation}
Rewriting the product of the two coordinate differentials in (\ref{e6}) in terms of the symmetric and non symmetric combinations, we get for the right side $\frac{1}{2}g_{\mu \nu}[(dx^\mu dx^\nu + dx^\nu dx^\mu )+ (dx^\mu dx^\nu - dx^\nu dx^\mu )]$, so that, we can write
\begin{equation}
g_{\mu \nu} = \eta_{\mu \nu} + kh_{\mu \nu}\label{e7}
\end{equation}
where the first term on the right side of (\ref{e7}) denotes the usual flat space time and the second term denotes the effect of the non commutativity, $k$ being a suitable constant.\\
It must be noted that if $l, \tau \to 0$ then equation (\ref{e7}) reduce to the usual formulation. From a physical point of view, if we are dealing with time and length scales much greater than the Compton wavelength, so that the order $0(l^2)$ terms can be neglected, then the usual commutative geometry works, with the usual derivates and more generally differential geometry. In that sense, and at such scales we can attribute the same meaning to coordinate differentials like $dx^\mu$. However this formulation breaks down at and inside the scale $(l, \tau)$. In what follows, in order to see the effect of the non commutative geometry, we will consider scales, near the minimum $(l,\tau)$ scale, and continue to use the concept of derivatives and differentials, incorporating the effects of departure from the commutative geometry.\\
The effect of the non commutative geometry is therefore to introduce a departure from flat space time, as can be seen from (\ref{e7}). Indeed, as is well known (Cf.ref.\cite{ohanian}), this is exactly as in the case of General Relativity and the second term on the right of (\ref{e7}) playing the role of the usual energy momentum tensor. However it must be borne in mind that we are now dealing with elementary particles. For an elementary particle, the material density vanishes outside its Compton wavelength and therefore also the minimum scale. On the other hand it shold be borne in mind that at and near the minimum scale itself we have the departure from the usual commutative geometry, as can be seen.\\
Infact remembering that the second term of the right side of (\ref{e7}) is small, this can straightaway be seen to lead to a linearized theory of General Relativity \cite{ohanian}. Exactly as in this reference we could now decue the General Relativistic relation
$$\partial_\lambda \partial^\lambda h^{\mu \nu} - (\partial_\lambda \partial^\nu h^{\mu \lambda} + \partial_\lambda \partial^\mu h^{\nu \lambda})$$
\begin{equation}
-\eta^{\mu \nu} \partial_\lambda \partial^\lambda h + \eta^{\mu \nu}\partial_\lambda \partial_\sigma h^{\lambda \sigma} = - k\bar T^{\mu \nu}\label{e8}
\end{equation}
It must be mentioned that the energy momentum type term on the right side of (\ref{e8}) arises due to the fact that the derivatives $\partial^\lambda$ and $\partial^\mu$ no longer commute and this leads to an additional contribution as can be verified from the left side of (\ref{e8}). To show this special origin of the right side term, we have used $\bar T$ instead of the usual $T$. More explicitely, it follows from the foregoing that (Cf.ref.\cite{bgsgrav})
\begin{equation}
\frac{\partial}{\partial x^\lambda} \frac{\partial}{\partial x^\mu} - \frac{\partial}{\partial x^\mu} \frac{\partial}{\partial x^\lambda} \mbox{goes over to} \frac{\partial}{\partial x^\lambda} \Gamma^\nu_{\mu \nu} - \frac{\partial}{\partial x^\mu} \Gamma^\nu_{\lambda \nu}\label{e9}
\end{equation}
Normally in conventional theory the right side of (\ref{e9}) would vanish. Let us designate this nonvanishing part on the right by
\begin{equation}
\frac{e}{c\hbar} F^{\mu \lambda}\label{e10}
\end{equation}
We have shown here that the non commutativity in momentum components leads to an effect that can be identified with electromagnetism and infact from expression (\ref{e10}) we have
\begin{equation}
A^\mu = \hbar \Gamma^{\mu \nu}_\nu\label{e11}
\end{equation}
where $A_\mu$ can be identified with the electromagnetic four potential (Cf.also ref.\cite{bgsgrav}).  To see this in the light of the usual guage invariant minimum coupling (Cf.ref.\cite{cu}), we start with the effect of an infinitessimal parallel displacement of a vector in this non commutative geometry,
\begin{equation}
\delta a^\sigma = - \Gamma^\sigma_{\mu \nu} a^\mu dx^\nu\label{e12}
\end{equation} 
As is well known, (\ref{e12}) represents the effect due to the curvature and non integrable nature of space - in a flat space, the right side would vanish. Considering the partial derivates with respect to the $\mu^{th}$ coordinate, this would mean that, due to (\ref{e12})
$$\frac{\partial a^\sigma}{\partial x^\mu} \to \frac{\partial a^\sigma}{\partial x^\mu} - \Gamma^\sigma_{\mu \nu} a^\nu$$
The second term on the right side can be written as:
$$-\Gamma^\lambda_{\mu \nu} g^\nu_\lambda a^\sigma = - \Gamma^\nu_{\mu \nu} a^\sigma$$
where we have utilised equation (\ref{e7}). That is we have
$$\frac{\partial}{\partial x^\mu} \to \frac{\partial}{\partial x^\mu} - \Gamma^\nu_{\mu \nu}$$
Comparison with (\ref{e11}) establishes the required identification.\\
It is quite remarkable that equation (\ref{e11}) is mathematically identical to Weyl's unified formulation, though this was not originally acceptable because of the adhoc insertion of the electromagnetic potential. Here in our case it is a consequence of the non commutative geometry (Cf.refs.\cite{cu} and \cite{bgsgrav} for a detailed discussion).\\
We can see this in greater detail as follows. The gravitational field equations can be written as \cite{ohanian}
\begin{equation}
D \phi^{\mu \nu} = - k\bar T^{\mu \nu}\label{e13}
\end{equation}
where
\begin{equation}
\phi^{\mu \nu} = h^{\mu \nu} - \frac{1}{2} \eta^{\mu \nu} h\label{e14}
\end{equation}
It also follows, if we use the usual guage and equation (\ref{e11}) that
\begin{equation}
\partial_\mu h^{\mu \nu} = A^\nu\label{e15}
\end{equation}
in this linearised theory.\\
Whence, remembering that we have (\ref{e7}), operating on both sides of equation (\ref{e13}) with $\partial_\mu$ we get Maxwell's equations of electromagnetism.\\
This is not surprising because as is well known if equation (\ref{e11}) holds as in the Weyl formulation, then in the absence of matter the general relativistic field equations (\ref{e8}) reduce to Maxwell equations \cite{berg}. In any case, all this provides a rationale for the fact that from (\ref{e13}) we get the equation for spin 2 gravitons (Cf.ref.\cite{ohanian}) while from the Maxwell equations, we have Spin 1 (vector) photons.
\section{Cosmology}
Based on the above model of fuzzy space time, we can deduce a cosmology, which infact correctly predicted against the tide an accelerating ever expanding universe. We use the fact that the Compton scale $l,\tau$ define minimum space time intervals as also the fact that particles are created out of the Zero Point Field or background Quantum Vaccum as discussed. Infact in such fluctuations, as is well known, given $N$ particles $\sqrt{N}$ particles are fluctuationally crated \cite{ijtp,newcosmos}. We consider as in the literature the pion to be a typical elementary particle, there being, as is well known $10^{80}$ such particlesin the universe.\\
In the following we will use $N$ as the sole cosmological parameter.\\
Equating the gravitational potential energy of the pion in a three dimensional isotropic sphere of pions of radius $R$, the radius of the universe, with the rest energy of the pion, we can deduce the well known relation \cite{nottale,hayakawa}
\begin{equation}
R \approx \frac{GM}{c^2}\label{e16}
\end{equation}
where $M$ can be obtained from the above.\\
We now use the fact that given $N$ particles, the fluctuation in the particle number is of the order $\sqrt{N}$ \cite{hayakawa,huang,ijmpa,ijtp,fqp,mg8} while a typical time interval for the fluctuations is $\sim h/mc^2$, the Compton time. We will come back to this point later. So we have
$$\frac{dN}{dt} = \frac{\sqrt{N}}{\tau}$$
whence on integration we get,
\begin{equation}
T = \frac{\hbar}{mc^2} \sqrt{N}\label{e17}
\end{equation}
We can easily verify that equation (\ref{e17}) is indeed satisfied where $T$ is the age of the universe. Next by differentiating (\ref{e16}) with respect to $t$ we get
\begin{equation}
\frac{dR}{dt} \approx HR\label{e18}
\end{equation}
where $H$ in (\ref{e18}) can be identified with the Hubble Constant, and using (\ref{e16}) is given by,
\begin{equation}
H = \frac{Gm^3c}{\hbar^2}\label{e19}
\end{equation}
Equation (\ref{e16}) and (\ref{e17}) show that that in this formulation, the correct mass, radius and age of the universe can be deduced given $N$ as the sole cosmological or large scale parameter Equation (\ref{e19}) can be written as
\begin{equation}
m \approx \left(\frac{H\hbar^2}{Gc}\right)^{\frac{1}{2}}\label{e20}
\end{equation}
Equation (\ref{e20}) has been empirically known as an ``accidental'' or ``mysterious'' relation. As observed by Weinberg \cite{weinberg}, this is unexplained: it relates a single cosmological parameter $H$ to constants from microphysics. We will touch upon this micro-macro nexus again. In our formulation, equation (\ref{e20}) is no longer a mysterious coincidence but rather a consequence.\\
As (\ref{e19}) and (\ref{e18}) are not exact equations but rather, order of magnitude relations, it follows that a small cosmological constant $\wedge$ is allowed such that
$$\wedge \leq 0 (H^2)$$
This is consistent with observatioins and shows that $\wedge$ is very very small - this has been a puzzle, the so called cosmological constant problem \cite{wein}. But it is explained here.\\
To proceed we observe that because of the fluctuation of $\sim \sqrt{N}$ (due to the ZPF), there is an excess electrical potential energy of the electron, which infact we have identified as its inertial energy. That is \cite{ijtp,hayakawa},
$$\sqrt{N} \epsilon^2 /R \approx mc^2$$
On using (\ref{e16}) in the above, we recover the well known Gravitation-electromagnetism ratio viz.,
\begin{equation}
e^2/Gm^2 \sim \sqrt{N} \approx 10^{40}\label{e21}
\end{equation}
or without using (\ref{e16}), we get, instead, the well known so called Eddington formula,
\begin{equation}
R = \sqrt{N} l\label{e22}
\end{equation}
Infact (\ref{e22}) is the spatial counterpart of (\ref{e17}). If we combine (\ref{e22}) and (\ref{e16}), we get,
\begin{equation}
\frac{Gm}{lc^2} = \frac{1}{\sqrt{N}} \propto T^{-1}\label{e23}
\end{equation}
where in (\ref{e23}), we have used (\ref{e17}). Following Dirac (cf. also \cite{melnikov} we get $G$ as the variable, rather than the quantities $m,l,c$ and $\hbar$ (which we will call microphysical constants) because of their central role in atomic (and sub atomic) physics.\\
Next if we use $G$ from (\ref{e23}) in (\ref{e19}), we can see that
\begin{equation}
H = \frac{c}{l} \frac{1}{\sqrt{N}}\label{e24}
\end{equation}
Thus apart from the fact that $H$ has the same inverse time dependance on $T$ as $G$, (\ref{e24}) shows that given the microphysical constants, and $N$, we can deduce the Hubble Constant also as from (\ref{e24}) or (\ref{e19}).\\
Using (\ref{e16}), we can now deduce that
\begin{equation}
\rho \approx \frac{m}{l^3} \frac{1}{\sqrt{N}}\label{e25}
\end{equation}
Next (\ref{e22}) and (\ref{e17}) give,
\begin{equation}
R = cT\label{e26}
\end{equation}
(\ref{e25}) and (\ref{e26}) are consistent with observation.\\
Finally, we observe that using $M,G \mbox{and}H$ from the above, we get
\begin{equation}
M = \frac{c^3}{GH}\label{e27}
\end{equation}
The relation (\ref{e27}) is required in the Friedman model of the expanding universe (and the Steady State model also).\\
The above model predicts an ever expanding and possibly accelerating universe whose density keeps decreasing. This seemed to go against the accepted idea that the density of the universe equalled the critical density required for closure.
\section{Issues and Ramifications}
i) The above cosmology exhibits a time variation of the gravitational constant of the form
\begin{equation}
G = \frac{\beta}{T}\label{e28}
\end{equation}
Indeed this is true in a few other schemes also, including Dirac's cosmology (Cf. \cite{narlikar,barrow,cu}). Interestingly it can be shown that such a time variation can explain the precession of the perihelion of Mercury (Cf.\cite{nc}). It can also provide an alternative explanation for dark matter and the bending of light while the Cosmic Microwave Background Radiation is also explained (Cf.\cite{cu}).\\
It is also possible to deduce the existence of gravitational waves given (\ref{e28}). To see this quickly let us consider the Poisson equation for the metric $g_{\mu \nu}$
\begin{equation}
\nabla^2 g_{\mu \nu} = G \rho u_\mu u_\nu\label{e29}
\end{equation}
The solution of (\ref{e29}) is given by
\begin{equation}
g_{\mu \nu} = G \int \frac{\rho u_\mu u_\nu}{|\vec r - \vec r'|} d^3 \vec r\label{e30}
\end{equation}
Indeed equations similar to (\ref{e29}) and (\ref{e30}) hold for the Newtonian gravitational potential also. If we use the second time derivative of $G$ from (\ref{e28}) in (\ref{e30}), along with (\ref{e29}), we can immediately obtain the D'alembertian wave equation for gravitational waves, instead of the Poisson equation:
$$D g_{\mu \nu} \approx 0$$
ii) Recently a small variation with time of the fine structure constant has been detected and reconfirmed by Webb and coworkers \cite{webb,webb2}. This observation is consistent with the above cosmology. We can see this as follows. We use an equation due to Kuhne \cite{kuhne}
\begin{equation}
\frac{\dot \alpha_z}{\alpha_z} = \alpha_z \frac{\dot H_z}{H_z},\label{e31}
\end{equation}
If we now use the fact that the cosmological constant $\wedge$ is given by
\begin{equation}
\wedge \leq 0 (H^2)\label{e32}
\end{equation}
as can be seen from (\ref{e18}), in (\ref{e31}), we get using (\ref{e32}),
\begin{equation}
\frac{\dot \alpha_z}{\alpha_z} = \beta H_z\label{e33}
\end{equation}
where $\beta < -\alpha_z < - 10^{-2}$.\\
Equation (\ref{e33}) can be shown to be the same as
\begin{equation}
\frac{\dot \alpha_z}{\alpha_z} \approx -1 \times 10^{-5} H_z,\label{e34}
\end{equation}
which is the same as Webb's result.\\
We give another derivation of (\ref{e34}) in the above context wherein, as the number of particles in the universe increases with time, we go from the Planck scale to the Compton scale.\\
This can be seen as follows: In equation (\ref{e21}), if the number of particles in the universe, $N = 1$, then the mass $m$ would be the Planck mass. In this case the classical Schwarzschild radius of the Planck mass would equal its Quantum Mechanical Compton wavelength. To put it another way, all the energy would be gravitational (Cf.\cite{cu} for details). However as the number of particles $N$ increases with time, according to (\ref{e17}), gravitation and electromagnetism get differentiated and we get (\ref{e21}) and the Compton scale.\\
It is known that the Compton length, due to zitterbewegung causes a correction to the electrostatic potential which an orbiting electron experiences, rather like the Darwin term \cite{bd}.\\
Infact we have
$$\langle \delta V \rangle = \langle V (\vec r + \delta \vec r)\rangle - V \langle (\vec r )\rangle$$
$$= \langle \delta r \frac{\partial V}{\partial r} + \frac{1}{2} \sum_{\imath j} \delta r, \delta r_j \frac{\partial^2 V}{\partial r_\imath \partial r_j}\rangle$$
\begin{equation}
\approx 0 (1) \delta r^2 \nabla^2 V\label{e35}
\end{equation}
Remembering that $V = e^2/r$ where $r \sim 10^{-8}cm$, from (\ref{e35}) it follows that if $\delta r \sim l$, the Compton wavelength then
\begin{equation}
\frac{\Delta \alpha}{\alpha} \sim 10^{-5}\label{e36}
\end{equation}
where $\Delta \alpha$ is the change in the fine structure constant from the early universe. (\ref{e36}) is an equivalent form of (\ref{e34}) (Cf.ref.\cite{kuhne}), and is the result originally obtained by Webb et al (Cf.refs.\cite{webb,webb2}).\\
iii) The latest observations of distaant supernovae referred to above indicate that the closure parameter $\Omega \leq 1$.\\
Remembering that $\Omega$ is given by \cite{ohanian}
$$\Omega = \frac{8 \pi G}{3H^2}\rho$$
we get therefrom on using 
$$\frac{H^2}{2G} R^3 = mN$$
which immediately leads to the mysterious Weinberg formula (\ref{e20}). Thus this is the balance between the cosmos at large and the micro cosmos.\\
iv) In General Relativity as well as in the Newtonian Theory, we have, without a cosmological constant
\begin{equation}
\ddot R = -\frac{4}{3} \pi G \rho R\label{e37}
\end{equation}
We remember that there is an uncertainity in time to the extent of the Compton time $\tau$, and also if we now use the fact that $G$ varies with time, \ref{e37}) becomes on using (\ref{e28}),
$$\ddot R = -\frac{4}{3} \pi G(t - \tau )\rho R$$
\begin{equation}
= - \frac{4}{3} \pi G \rho R + \frac{4}{3} \pi \rho R \left(\frac{\tau}{t}\right) G\label{e38}
\end{equation}
Remembering that at any point of time, the age of the universe, that is $t$
itself is given by (\ref{e17}), we can see from (\ref{e38}) that this effect of time variation of $G$; which again is due to the background Zero Point Field is the same as an additional density, the vacuum density given by
\begin{equation}
\rho_{vac} = \frac{\rho}{\sqrt{N}}\label{e39}
\end{equation}
This term in (\ref{e38}) is also equivalent to the presence of a cosmological constant $\wedge$ as discussed above. On the other hand, we know independently that the presence of a vacuum field leads to a cosmological constant given by (Cf.ref.\cite{cu} and references therein)
\begin{equation}
\wedge = G \rho_{vac}\label{e40}
\end{equation}
Equation (\ref{e40}) is pleasingly in agreement with (\ref{e38}) and (\ref{e39}) that is, to the density of the fluctautionally created particles in the vacuum. In other words quantitatively we have reconfirmed that it is the background Zero Point Field that manifests itself as the cosmological constant described. This also gives as pointed out an explanation for the so called cosmological constant problem \cite{wein} viz., why is the cosmological constant so small?\\
v) In the above cosmology of fluctuations, our starting point was the creation of $\sqrt{N}$ particles within the minimum time interval, a typical elementary particle Compton time $\tau$. A rationale for this, very much in the spirit of the condensation of particles from a background Zero Point Field as discussed at the beginning has also been obtained recently in terms of a broken symmetry phase transition from the Zero Point Field or Quantum Vacuum. In this case, particles are like the Benard cells which form in fluids, as a result of a phase transition. While some of the particles or cells may revert to the Zero Point Field, on the whole there is a creation of these particles. If the average time for the creation of one of these particles or cells is $\tau$, then at any point of time where there are $N$ such particles, the time elapsed, in our case the age of the universe, would be given by (\ref{e17}) (Cf.\cite{bgsnature}). While this is not exactly the Big Bang scenario, there is nevertheless a rapid creation of matter from the background Quantum Vacuum or Zero Point Field. Thus half the matter of the universe would have been created within a fraction of a second.\\
In any case when $\tau \to 0$, we recover the Big Bang scenario with a singular creation of matter, while when $\tau \to$ Planck time we recover the Prigogine Cosmology (Cf.\cite{cu} for details). However in neither of these two limits we can deduce all the above consistent with observation relations.\\
vi) The discrete spacetime effects, it was argued, lead to a deviation from the relativistic energy-momentum formula \cite{xx,cu}, which again is an $0(l^2)$ effect. This can be seen if we invoke Wilson's Lattice theory to get \cite{mont},
\begin{equation}
E^2 = m^2 + p^2 + \lambda l^2, \quad \lambda = -\frac{1}{3}p^4,\label{e41}
\end{equation}
Effectively, this is a correction on the velocity of light, $c^2 \to c^2(1 + \epsilon)$. Using (\ref{e41}), the correction term $\epsilon$ is in agreement wlith the phenomenological constraint on it given by Glashow and Coleman \cite{cole}. It also agrees with the empirical form of Lorentz Symmetry Violation \cite{jacob}. All this explains the High Energy Cosmic Ray puzzle, viz., the observed violation of the relativistic GZK cut off for these Ultra High Energy Cosmic Rays \cite{olin,caro,naga}.\\
vii) The fuzzy spacetime considerations, lead to a short range force (or massive photon force) \cite{cu,xphys,bgsrf}, some evidence for which seems to be available already, as described in the references.\\ \\
{\large {\bf ACKNOWLEDGEMENT}}\\ \\
\noindent The author is thankful to Prof. H. Kr\"{o}ger of the University of Laval, Canada, for his hospitality and useful discussions.

\end{document}